\documentclass[sigconf]{acmart}

\AtBeginDocument{%
  \providecommand\BibTeX{{%
    \normalfont B\kern-0.5em{\scshape i\kern-0.25em b}\kern-0.8em\TeX}}}

\setcopyright{iw3c2w3}
\copyrightyear{2022}
\acmYear{2022}
\acmConference[DIS '22]{Designing Interactive Systems Conference}{June 13--17, 2022}{Virtual Event, Australia}
\acmBooktitle{Designing Interactive Systems Conference (DIS '22), June 13--17, 2022, Virtual Event, Australia}\acmDOI{10.1145/3532106.3533496}
\acmISBN{978-1-4503-9358-4/22/06}

\newtoggle{comments}
\toggletrue{comments}

\iftoggle{comments} {
  \newcommand {\blair}[1]{{\color{purple}\bf{blair: #1}\normalfont}}
}{
  \newcommand {\nadya}[1]{}
  \newcommand {\blair}[1]{}
 }
\usepackage{graphics}      %
\usepackage{listings}
\usepackage{stfloats}

\definecolor{background}{HTML}{EEEEEE}
\definecolor{p5}{RGB}{11,124,169}
\definecolor{p5math}{RGB}{213,40,137}

\lstdefinelanguage{json}{
    basicstyle=\normalfont\ttfamily,
    stepnumber=1,
    showstringspaces=false,
    breaklines=true,
    frame=lines,
    backgroundcolor=\color{background},
    string=[s]{"}{"},
    stringstyle=\color{darkgray}\bfseries,
    comment=[l]{:},
    commentstyle=\normalfont\ttfamily
}

\lstdefinelanguage{javascript}{
    basicstyle=\fontsize{8}{10}\selectfont\ttfamily,
    stepnumber=1,
    showstringspaces=false,
    breaklines=true,
    frame=lines,
    backgroundcolor=\color{background},
    comment=[l]{//},
    morecomment=[s]{/*}{*/},
    commentstyle=\color{gray}\ttfamily,
    stringstyle=\color{red}\ttfamily,
    morestring=[b]',
    morestring=[b]",
    identifierstyle=\color{black},
    keywords=[1]{typeof, new, true, false, catch, function, return, null, catch, switch, var, if, in, while, for, do, else, case, break, let},
    keywordstyle= [1]{\color{darkgray}\bfseries},
    keywords = [3]{PI, TWO_PI},
    keywordstyle = [3]{\color{p5math}},
    keywords = [4]{map, sin, cos, maxm floor},
    keywordstyle = [4]{\color{p5}\bfseries},
    ndkeywords={class, export, boolean, throw, implements, import, this, fab, moveExtrude, moveRetract, setStartingAcceleration, setMaxAcceleration, setJerk, autoHome, setNozzleTemp, setBedTemp},
    ndkeywordstyle=\color{p5},
    sensitive=false
}
\begin{document}

\title{\texttt{p5.fab}: Direct Control of Digital Fabrication Machines from a Creative Coding Environment}

\renewcommand{\shorttitle}{\texttt{p5.fab}}
\author{Blair Subbaraman}
\email{b1air@uw.edu}
\affiliation{%
  \institution{University of Washington}
  \city{Seattle}
  \state{Washington}
  \country{USA}
  \postcode{91895}
}
\author{Nadya Peek}
\email{nadya@uw.edu}
\affiliation{%
  \institution{University of Washington}
  \city{Seattle}
  \state{Washington}
  \country{USA}
  \postcode{98195}
}

\begin{abstract} 
Machine settings and tuning are critical for digital fabrication outcomes. 
However, exploring these parameters is non-trivial.
We seek to enable exploration of the full design space of digital fabrication. 
To identify where we might intervene, we studied how practitioners approach 3D printing.
We found that beyond using CAD/CAM, they create bespoke routines and workflows to explore interdependent material and machine settings.
We seek to provide a system that supports this workflow development. We identified design goals around material exploration, fine-tuned control, and iteration.
Based on these, we present \texttt{p5.fab}, a system for controlling digital fabrication machines from the creative coding environment p5.js.
We demonstrate \texttt{p5.fab} with examples of 3D prints that cannot be made with traditional 3D printing software.
We evaluate \texttt{p5.fab} in workshops and find that it encourages novel printing workflows and artifacts.
Finally, we discuss implications for future digital fabrication systems.
\end{abstract}



\begin{teaserfigure}
  \includegraphics[width=\textwidth]{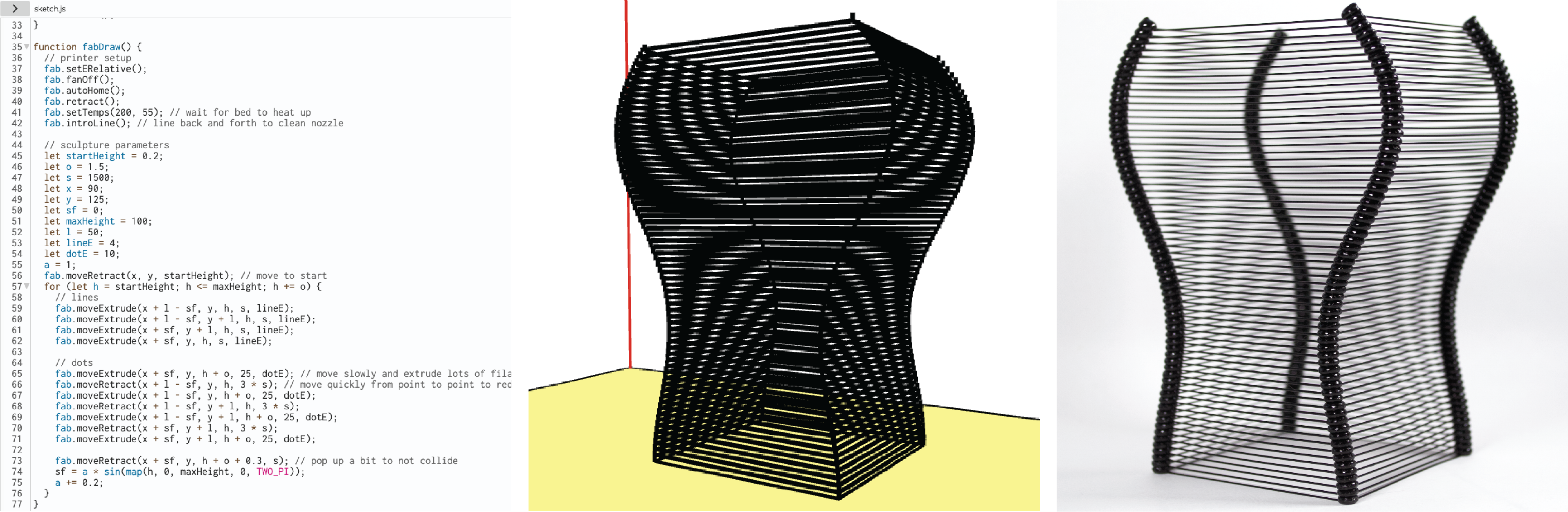}
  \caption{Our system, \texttt{p5.fab}, controls digital fabrication machines from the creative coding environment p5.js. We programatically generate toolpaths using our library (left) which we can visualize in our user interface with a render command (center). We stream these commands to a desktop FFF 3D printer for fabrication; a photo of the resulting print is shown (right).}
  \Description{}
  \label{fig:teaser}
\end{teaserfigure}

\maketitle

\section{Introduction}
In the late 1960s, artist Charles Csuri developed a new method for making sculpture with undulating, mathematically described surfaces \cite{csuri_technique_1971}. 
After visualizing a surface with the help of a mainframe computer, a program produced punched tape representing the sculpture’s coordinate data in a format suitable for a 3-axis, continuous path, computer-numerically-controlled (CNC) milling machine. 
In the artist comments for \textit{Numeric Control} (1968)—one of the first artworks created using a CNC milling machine—Csuri wrote that ``While the device was capable of making a smooth surface, I decided it was best to leave the tools marks for the paths'' \cite{csuri_numeric_1968}. 
Csuri used the ridged grain of the toolpath's marks in his final work, retaining creative control not just through the overall form of his piece, but also through details of the surface finish that were defined by the fabrication method.

\begin{figure*}
    \centering
    \includegraphics[width=1\linewidth]{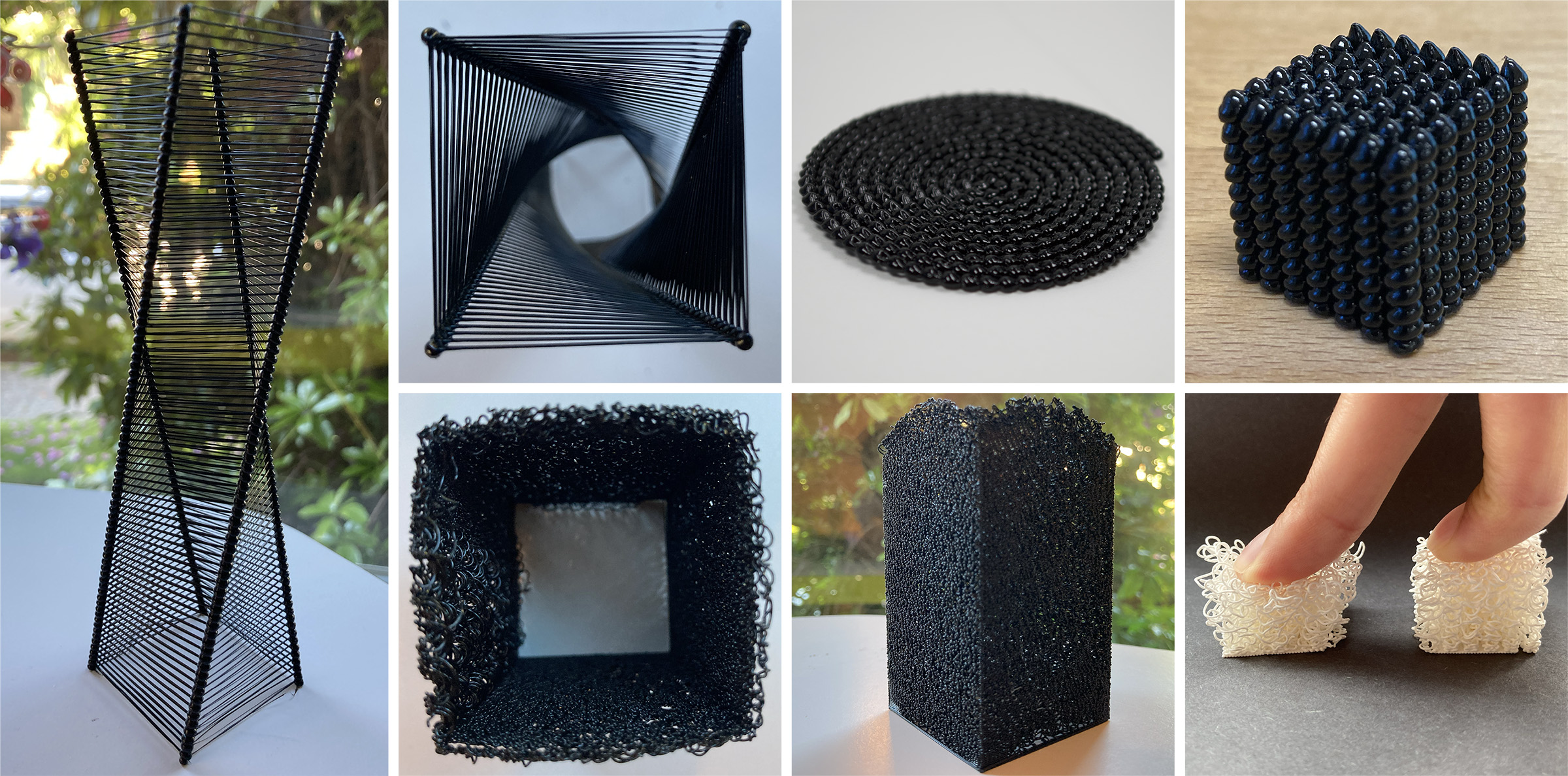}
    \caption{FFF 3D prints made using \texttt{p5.fab}. These prints would not be possible with off-the-shelf slicer software. Clockwise from left: bridged filament is anchored with one-shot extruded dots. Bridged filament top view. A 2D spiral of dots. A 3D dot print. Foam prints with varying density in compression. Foam print side view. Foam print top view.}
    \label{fig:experimentalprints}
\end{figure*}

Csuri’s technique for sculpture took the physical features of CNC milling toolpaths into account, using them creatively in a way that often isn't accounted for in contemporary digital fabrication workflows.
Contemporary CAM software for digital fabrication machines (such as CNC mills, laser cutters, and 3D printers) is designed for users to optimize toolpaths such that they will create a physical part that is as faithful to the CAD model as possible.
For example, in fused filament fabrication (FFF, synonymous with FDM) 3D printing, standard CAM settings equate bumpier surface finishes with lower quality parts and smoother ones with higher quality.
While there have been considerable efforts in HCI to enable alternative control of digital fabrication machines \cite[e.g.][]{willis_interactive_2010, peng_roma_2018, zoran_freed_2013}, this work runs counter to mainstream practice. 
We identify an opportunity to better enable the exploration of digital form together with the real-world materiality of digital fabrication.

We believe that the steps of digital fabrication workflows are intertwined; we agree they do not follow ``a canonical workflow that proceeds rigidly through several consecutive stages: a design intent, a digital representation, machine instructions, and a final product'' \cite{twigg-smith_tools_2021}.
Digital fabrication involves not only machines and software, but also physical materials, tools, and the manner by which the practitioner stitches these components together. 
Changing the choice of materials or physical tools (e.g., end mills, nozzles, lenses) impacts outcomes as much as changing software settings. Moreover, the state of the equipment determines digital fabrication outcomes.
Software settings can be repeated consistently and exactly; the condition of the machine, material, and environment cannot. 
We consequently identify \emph{calbration and tuning} of machine settings and parameters as a critical part of digital fabrication practice, mandating an intimate collaboration between human, machine, and material. 
 
 We therefore ask:
 \emph{How can we develop systems for digital fabrication than enable iterative exploration of the full range of parameters, from digital models to selected materials to machine tuning?}
 In doing so, we invest in the idea that ``digital fabrication can be more than a series of steps, that materials have agency alongside a user’s goals, and that opportunities for creative exploration are more important
than seamless control'' \cite{twigg-smith_tools_2021}. 

We selected FFF 3D printers for our study due to their widespread adoption as digital fabrication machines.
To better understand current 3D printing practices, especially how practitioners tie printing outcomes to machine state and settings, we conducted a formative study. 
We use the findings from 10 semi-structured interviews and 92 survey responses from participants recruited from online 3D printer debugging and repair communities to conceptualize themes. 
This inquiry sheds light on the ways in which practitioners develop their own routines to understand the feedback loops in their digital fabrication workflows.
In our formative study, we found that practitioners developed their own ad-hoc tuning routines, 
negotiated digital and physical settings,
and explored CAM tools creatively rather than as a way to optimize. 
Overall, we observed that practitioners developed a craft sensibility for 3D printing but that they were often hampered by the software they used.

Using insights from our formative study, we determined the following system design goals: (1) promoting practitioner agency through open-ended workflow exploration, (2) facilitating iterative material interaction, and (3) integrating digital design intent with physical making through toolpath control. 

Using these design goals, we present \texttt{p5.fab}: a system which enables direct control of machines from the popular creative coding environment, p5.js. \texttt{p5.fab} includes of a library for programmatic control of machines within p5.js, a user interface to visualize toolpaths, and GUI elements for sending commands and receiving positions. Compared to the roundabout control offered by popular CAD and CAM software, we offer low-level control over machine toolpaths. These commands can be sent directly and immediately to the machine. We demonstrate our system through a series of example workflows which center material exploration. 

We evaluate \texttt{p5.fab} in two workshops, first with three professional artists who use code as an expressive medium, then with three makers experienced in 3D printing.
We find that \texttt{p5.fab} offers our participants a familiar entry-point into digital fabrication which takes advantage of their artistic and craft sensibilities. 
We detail evaluation insights, examining how participants work with the system towards creative goals. We conclude with implications for future systems which incorporate direct machine control.

Our contributions are:
\vspace{-4pt}
\begin{itemize}
    \item Insights into 3D printing practice from our formative study
    \item The interactive printing environment \texttt{p5.fab}
    \item An evaluation of our system through workshops.
\end{itemize}

\section{Background \& Related Work}
Our research contributes to both HCI systems research in digital fabrication and HCI inquiry into situated machine practice.
In this section, we provide an overview of prior work detailing how we build upon and distinguish from other systems.

\subsection{Digital Fabrication Systems Research}
CAD software has steep learning curves and built-in assumptions about expected use \cite{hudson_understanding_2016}. HCI researchers have made a significant effort to decrease the barrier to entry, including through gesture based 3D modeling \cite{weichel_mixfab_2014, gannon_tactum_2015}, programmatic 3D modeling \cite{yeh_craftml_2018, maleki_liveness_2014}, and remixing of online models \cite{hofmann_greater_2018}. 
These projects generally target newcomers to fabrication and assume a traditional divide between digital design and machine execution. Other research trajectories have dissolved these boundaries. Hybrid craft has incorporated craft experts in digital fabrication \cite{zoran_hybrid_2013, zoran_hybrid_2014, jacobs_hybrid_2015} and alleviated the requisite manual skill through smart tools \cite{tian_matchsticks_2018, tian_turn-by-wire_2019, zoran_freed_2013}. One of our goals is to engage creative coders in 3D printing, who bring with them their own craft programming expertise \cite{lingel_its_2014}. However, the design of our system is driven by qualitative insights into 3D printing practice. Accordingly, our system is not solely concerned with engaging newcomers but also reinvests in existing practitioners.

Grounding our contribution in digital fabrication practice, our work is aligned with interactive fabrication systems which make machine interactions more like craft processes \cite{willis_interactive_2010, kim_compositional_2018, peng_roma_2018}. While these works often involve real-time interactions, our system retains the precision of asynchronous design. 
Two projects most similar to our work also negotiate interactive fabrication and creative practice. LINC \cite{li_direct_2017} is a sketch-based tool to control CNC routers which allows artists to take advantage of direct control for asynchronous or real-time authoring. LINC begins with manual sketching practice to generate toolpaths; \texttt{p5.fab} offers programmatic control to cater to the precise nature of tuning work. \citet{fossdal_interactive_2021} contribute software to control a custom digital fabrication machine alongside an evaluation with a professional artist. Similarly focused on enabling toolpath control, their system centers expert practitioners by augmenting their CAD working environment, thereby relying on their expertise in CAD. Our work differs in that \texttt{p5.fab} is focused on interfaces to control existing machines by drawing connections between understudied sites of existing fabrication practice, namely, tuning \& calibration, to other modes of digital practice, namely, creative code.

CAM software for 3D printing takes the form of `slicers' which generate machine readable G-code from geometric models. Related work has achieved functional and expressive results by manipulating, extending, and creating new slicer software:  \textit{3D Printed Fabric} \cite{takahashi_3d_2019} and \textit{DefeXtiles} \cite{forman_defextiles_2020} leverage gap defects in the printing process to create woven structures, and \textit{Furbication} \cite{laput_3d_2015} exploits the stringing of filament to produce hair-like structures. 
\textit{Expressive FDM} \cite{takahashi_expressive_2017} contributes quantitative models of common 3D printing errors and appropriate this behaviour to achieve novel aesthetic results. Our system connects research into FFF printing parameters with direct control to omit the slicer entirely. 
Unlike prior work, we do not contribute an end-to-end workflow for making a specific type of object. Rather, we appeal to the craft nature of 3D printing to enable practitioners' exploration of machine and material behaviour. In this way, we draw similarities to artistic practice such as LIA's \textit{Filament Sculptures} \cite{lia_filament_2021}. 

Beyond slicing software, there are also approaches that involved direct G-Code editing and encapsulation. \citet{koda_direct_2017} have investigated direct G-Code manipulation to attain higher machine precision; in their 3D material "weaving", G-Code editing is a post-processing step preceded by conventional slicing. This follows functionality built-in to many CAM softwares to tweak G-Code using arbitrary scripts \cite{noauthor_slic3r_nodate}. A more radical shift is proposed by FullControl GCode Designer, which permits carefully designing toolpath parameters in Microsoft Excel. \cite{gleadall_fullcontrol_2021}. \citet{10.1145/3490149.3501312} also enact programmatic control from Python. Like \texttt{p5.fab}, these systems replace the slicer. With \texttt{p5.fab}, we have a particular focus on material exploration. To this end, we incorporate direct control of the machine to interpret material output before designing an entire artifact. We enact machine control from within a general-purpose programming language to encourage generative possibilities alongside custom UI elements. We believe this allows for quicker iteration and therefore more experimentation. 
Furthermore, our system supports commands beyond G-code, for example, it can also be used with the commercial Axidraw pen plotter's \cite{axidraw} interactive mode.

\subsection{3D Printing Practice \& Maintenance}
Moments of breakdown in our individual and collective interactions with technology can serve as new analytical starting points \cite{10.1145/2598784.2602775, suchman2007human, orr_talking_1996}. Working from rich ethnographic traditions studying maintenance and repair in HCI, our work privileges the experiences of 3D printing practitioners who share their experiences with us. Maintenance involves a wide array of activities. The participants in our formative study explain that the precise tuning of material and machine parameters comprises an important part of keeping printers in working condition. We believe the details of such calibration \& tuning processes can help clarify relationships between materials, machines, and operators to inform systems design. In the context of contemporary digital fabrication, we note that many machine users are now responsible for maintenance historically performed by expert technicians. We are interested in how direct machine control can support gaining the tacit knowledge they need to perform this work.

While our work engages with online 3D printer repair communities, we focus on individual relationships with machines. In \emph{Producing Printability}, \citet{dew_producing_2019} study 3D printing in practice to investigate what makes a design `printable'. They propose a new set of priorities for 3D printing design tools orthogonal to many HCI systems contributions which focus on print efficiency. Drawing on the concept of articulation work, or ``the work of fitting people and tasks into a broader process``, they argue for tools which support the messy realities of 3D printing practice rather than ones which ``smooth them over''. Related work by \citet{landwehr_sydow_machine_2020} contributes the concept of \textit{machine sensibility} to describe practitioner's ability to assess printability, tactfully intervene, and interpret the resultant print. Our study of 3D printing practice and subsequent design goals respond to these findings.
\section{Identifying System Design Goals}

To contribute systems which enable interactive exploration of the digital fabrication design space, we first seek to identify concrete system design goals.
Calibration and tuning work requires practitioners to continuously interpret material output. 
Therefore, we believe understanding how practitioners approach this work will give us key insights into how different choices impact fabrication outcomes.
Our goal is not to engineer cure-all solutions. Building on previous work, we are cognizant of how automating printing processes might unintentionally limit practitioners' control over fabrication outcomes \cite{dew_producing_2019, 10.1145/2598784.2602775}. Automation requires nuanced understanding of low-level parameters. Our design goals suggest systems which build up this understanding. From our themes, we draw out process-level insights to see where interactive fabrication systems can productively intervene and expand current practice.
In this section, we elaborate on our formative study and the design implications for our system.

\subsection{Formative Study Methods}
We recruited interview participants from two active online communities for troubleshooting 3D prints and printers. The subreddit r/FixMyPrint and the PrintEverything Discord are communities of 67k and 2k members respectively. We conducted 10 semi-structured interviews over the course of three months, with 7 participants (referred to as P1-P7) recruited from r/FixMyPrint and 3 participants (P8-P10) from the PrintEverything Discord. All participants personally owned and managed at least one FFF 3D printer. Interviews were conducted using video conferencing software and lasted 45 minutes on average. 

We transcribed interviews and conducted analysis using a modified grounded theory approach with inductive open coding to develop a codebook and themes over six months. To further dilate these codes, we made use of a memoing process to produce text that worked across the code set \cite{emerson_writing_nodate}. To account for the breadth of practice, we subsequently distributed a survey through a number of channels including Reddit, Discord, and Twitter. Survey questions were targeted to build upon themes generated from interviews and were cross-coded accordingly. We received 92 responses over two weeks. In a final meeting, we synthesized the findings from the survey and interview data. From each theme, we derive a design goal for a system to support calibration \& tuning work.

\subsection{Themes}
We conceptualized three main themes that we believe characterize current calibration and tuning practices. Practitioners we spoke with  develop ad-hoc routines in response to recurring problems. These routines bring practioners into intimate arrangement with machine and material. As the interlocutor between digital representation and physical object, CAM emerges as a contested site of practice. 

\subsubsection{Practitioners develop ad-hoc routines in response to recurring problems} While all participants encountered similar problems, we found that exactly how they chose to remedy the problem differed for each person. 
We reason that the divergent approaches we observed partially stem from the complexity and interdependence of parameters.
For example, all interview participants flag bed adhesion as a significant recurring issue, defined by P2 as a problem wherein the \textit{``print fails because the prints don't stick to the bed''}. Participants describe how they could address bed adhesion issues from different angles, such as physical maintenance like cleaning and leveling the build plate, digital changes like remodeling the object, editing CAM settings such as extrusion multipliers and speed attributes, or machine investigations into extrusion issues and mechanical checks. 
However, participants reported that while similar problems can be addressed with very different approaches, they relied on intuition they had built up from their own prior experience.

\subsubsection{Calibration \& Tuning Bring Practitioners into Intimate Arrangement with Machine \& Material} Further destabilizing the view that calibration \& tuning are a rote series of steps, practitioners we spoke with emphasized the importance of embodied knowledge. Through regular use of the printer, practitioners attune their vision to the subtle signals provided by the printing process. P4 explains that \textit{``you can usually tell when you've been doing it for a while... if the layer’s going down right. If it's squished just enough, if it's not squished enough.''}
The qualitative notion of ``squish'' refers to how tightly the nozzle is pressed against the bed as it extrudes filament. 
It is a function of many variables: the printer model, the type of filament, the curvature of the build plate, the temperature of the bed, nozzle diameter and state, the room the printer is in, and more might all be relevant. Practitioners are consequently drawn into intimate arrangement with the machine and material to monitor deposition. 
P2 will watch the first layer of their print and manually ``\textit{baby step}'' the adjustment knobs up \& down accordingly.
P9, as well as several survey respondents, will print a series of calibration files and fine tune both physical knobs and CAM settings based on the results. 
These routines enable practitioners to iteratively reconcile the digital with the physical world. The iteration is guided by material observation rather than achieving a design intent.

\subsubsection{CAM Tools Negotiate the Digital-Physical Divide} CAM software is where digital representations are connected with important physical variables like speed, acceleration, and temperature.
Participants explain that an important site of practice is in tuning these settings, also known as developing slicer ``profiles''. These profiles negotiate trade-offs inherent to the printing process. For example, P8 uses a slower print speed as they mostly make functional parts and appreciate the dimensional precision; P2 on the other hand runs `boundary tests' to operate their machines at their limits.  
 While print speed is just one example, slicers offer control over many parameters. Participants explain that it is often difficult to discern what effect changing these settings have. P1 had a clog that plagued their print quality for half a year. They observed that their printer was under-extruding, meaning less filament was exiting the nozzle than needed. However, they note that their printer never printed perfectly, and so the underlying issue went unnoticed for some time. Instead, they would compensate by changing slicer settings. While changing slicer parameters offered a quick fix, quality quickly deteriorated again. 
It was only when the pneumatic tube connector on the hot end popped out that they disassembled the hot end and found a clog. 
In the end, slicer settings obscured the problem. Because slicer settings are disconnected from machine execution, it can promote unhelpful troubleshooting estranged from physical context.

\subsection{Design Goals}
Based on these themes, we developed three corresponding design goals for our system. While we consider automated solutions, a key insight from our formative study is that calibration \& tuning requires in-depth understanding of interdependent parameters. This understanding is a prerequisite to automation. Consequently, our design goals seek facilitate exploration of low-level settings.

\subsubsection{Shift from Enumeration to Exploration} 
Participants' stories demonstrate that calibration \& tuning routines are workflows in and of themselves, developed to bring print quality into alignment with expectations. 
The varied nature of printer problems and practitioners' solutions suggest against providing a cure-all end-to-end system. We reject the option to develop auto-calibration procedures which attempt to address contingent trade-offs between digital models and physical parameters, as these solutions neglect the highly interdependent nature of the problems. Instead, we seek to support the work which practitioners are already engaged in to better facilitate exploration of interdependent settings.

We observe that lightweight G-code files are successful in scaffolding some procedures such as moving the nozzle around the work envelope for manual bed leveling. Support tools can take a step further by recognizing that these settings are physically coupled \cite{agarwala_structural_1996}; for example, increasing bed temperature to promote adhesion results in thermal expansion of the build plate which can void manual bed leveling. 
While G-code offers control over these various parameters, it is cumbersome to customize. Exposing machine commands in a manner more legible to humans offers one method to consolidate disparate settings. This design goal suggests a low-level intermediate representation which provides machine control outside the context of printing a model.

\subsubsection{Embed Opportunity for Tacit Knowledge Through Iterative Material Exploration}
As opposed to traditional 3D print workflows wherein design intent is expressed through CAD software, calibration routines take place in the physical world. It is ultimately the material outcome which informs the practitioner how to proceed. This process challenges conventional notions of the printer as an output device; maintenance is fundamentally conversational. Rather than automate away these critical activities, we aim to help practitioners develop their embodied expertise.  Therefore, our system design should permit commands to be sent to the machine to iteratively observe material outcomes. Low-level representations combined with direct control promotes turn-taking between operator and machine, encouraging small unit tests to explore material behavior. 

\subsubsection{Integrate Digital Design \& Physical Making through Toolpath Control}
 While slicers offer control over many parameters, we observe that they promote a monolithic approach to tuning settings detached from material output. Since editing a model requires reverting back to CAD software, slicers further overlook the relationship between geometry and physical parameters. Slicer toolpath calculations are sophisticated in their ability to optimize print time. However, they force practitioners to reason about static digital models geometry rather than dynamic material behavior. It should be straightforward to discern what effect changing the value of a parameter has on a print. Our system should grant continuous control over motion and material settings.

\begin{figure}[h!]
    \centering
    \includegraphics[width=1\linewidth]{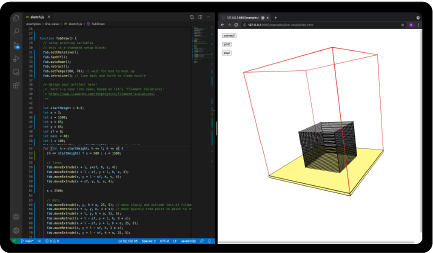}
    \caption{p5.fab interface. Toolpaths generated in code (left) are visualized in the user interface (right) then streamed to the machine over serial. %
    }
    \label{fig:systemoverview}
\end{figure}

\section{Programmatic Direct Control of 3D Printers}
In line with our three design goals, we develop a system which encourages open-ended workflow exploration, privileges iterative material interaction, and grants control over machine toolpaths.

We present \texttt{p5.fab}\footnote{Our source code is available online at \url{https://github.com/machineagency/p5.fab}.}, a system for programmatic direct control of digital fabrication machines from the creative coding environment p5.js. In doing so, \texttt{p5.fab} avoids jumping between different software and file formats, instead prioritizing rapid creative exploration.

\begin{figure*}[htbp]
\begin{minipage}{0.5\linewidth}
\begin{lstlisting}[language=javascript]
/* Define Lissajous parameters
  - A & B set the width & height
  - a & b change the number of lobes
  - delta rotates the curve
  - step discretizes the curve 
  - z is the first layer height */
let A = 100; 
let B = 100;
let a = 5;
let b = 4;
let delta = PI/2;
let step = TWO_PI/200;
let z = 0.2;

for (t = 0; t <= TWO_PI; t += step) {
  x = A * sin(a * t + delta);
  y = B * sin(b * t);
  /* moveExtrude calculates default
     velocity and extrusion amounts
     if called without explicit
     values. */
  fab.moveExtrude(fab.maxX / 2 + x,
                  fab.maxY / 2 + y,
                  z);
}
\end{lstlisting}
\end{minipage}
\begin{minipage}{0.45\linewidth}
\includegraphics[width=\linewidth]{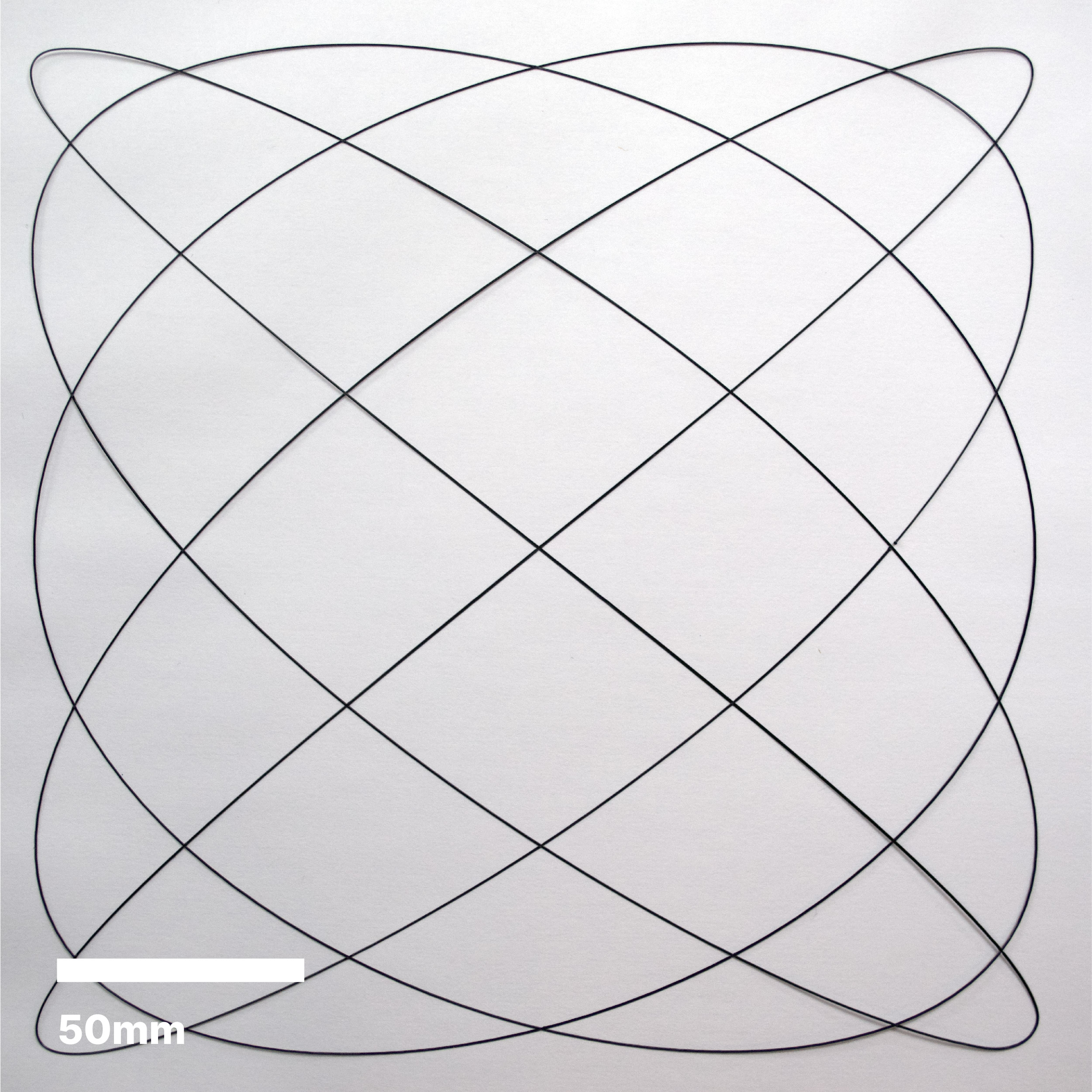}
\end{minipage}
\caption{We programmatically define a toolpath using the equations for a Lissajous curve (left), where we quickly iterate using variables for geometry (e.g. size, number of lobes) and variables for material properties (e.g. speed, extrusion rate). A photograph of the printed result is shown on the right.}
\label{fig:lissajous}
\end{figure*}

We develop our system contribution as a library for p5.js, a popular open-source JavaScript library for creative coding \cite{noauthor_p5js_2021}. p5.js (or p5 for short) uses the metaphor of a sketchbook, attempting to make drawing with code as intuitive as drawing with paper and pen. Previous research has argued for pairing art production with tool production to foster broader participation in creative system development \cite{li_what_2021}. p5 has over 1.5 million users, known also as creative coders \cite{noauthor_p5js_2021-1}. Our tool is therefore well positioned to be used by this large, active, and creative community.

\subsection{Using \texttt{p5.fab}}
\label{using-p5fab}
Before providing \texttt{p5.fab}'s implementation details, we describe how to use \texttt{p5.fab} to make the prints in Figures \ref{fig:teaser} and \ref{fig:experimentalprints}.

To make the bridged filament print shown in Figure \ref{fig:experimentalprints}, we take the following steps: We connect our computer to a 3D printer using a USB cable and open a text editor. As this is a type of filament we have never used before, we want to quickly test how long a bridge we can make by horizontally connecting two points without support material. We program a series of test prints to make longer and longer bridges. The longest bridges droop in the middle. Therefore, we try higher speeds until we can reliably print our max bridge length. We also experiment with thicker lines (made by extruding more filament over the same distance), which due to increased mass need to be printed at a higher speed. After finding a suitable new speed, we place the lines of code creating these toolpaths in a \lstinline{for} loop, stacking them to assemble larger structures like in Figure \ref{fig:teaser}. This workflow is not possible with off-the-shelf CAD/CAM systems.

 To provide a better sense of what goes into a \texttt{p5.fab} code sketch, Figure \ref{fig:lissajous} shows the code that goes into a simple print. In a few lines, we can specify the equation of a Lissajous curve. This curve is immediately previewed on the \texttt{p5} canvas. We can change our geometry by editing the parameters. From our curve, we generate a set of coordinates for the extruder to pass through. We can either save this code or stream the toolpaths to the printer. 

In the code in Figure \ref{fig:lissajous}, the base \lstinline{fab} object represents the machine. A \lstinline{fab} object is instantiated with machine data including work envelope dimensions, nozzle radius, and filament radius. We can therefore easily center the print on the bed using the \lstinline{maxX} and \lstinline{maxY} properties. When called without explicit values, the \lstinline{moveExtrude} method will move to the specified coordinate at a default speed, and calculate quantity of material deposition assuming single width extrusion. By expressing the physical object and the machine parameters which made them as programmatic constructs, we hope to make the programs more easily readable across machines and setups.

\begin{figure}[htbp]
\begin{lstlisting}[language=javascript]
/* Define velocity painting parameters */
let speedA = 30;
let speedB = 20;
let cubeLen = 20;
let checkerLen = 5;
let layerHeight = 0.2;

/* Create a checkered surface finish by alternating
  between slower and faster regions */
for (z = 0; z <= cubeLen; z += layerHeight) {
  region = (floor(z/checkerLen) %
  for (x = xStart; x <= xEnd; x += checkerLen) {
    speed = region ? speedA : speedB;
    fab.moveExtrude(x, y, z, speed);
    region = !region
  }
  /* Print other sides of the cube without 
     any velocity painting */ 
  fab.moveExtrude(xStart+cubeLen, y+cubeLen, z);
  fab.moveExtrude(xStart, y + cubeLen, z);
  fab.moveExtrude(x, y, z);
}
\end{lstlisting}
\includegraphics[width=1\linewidth]{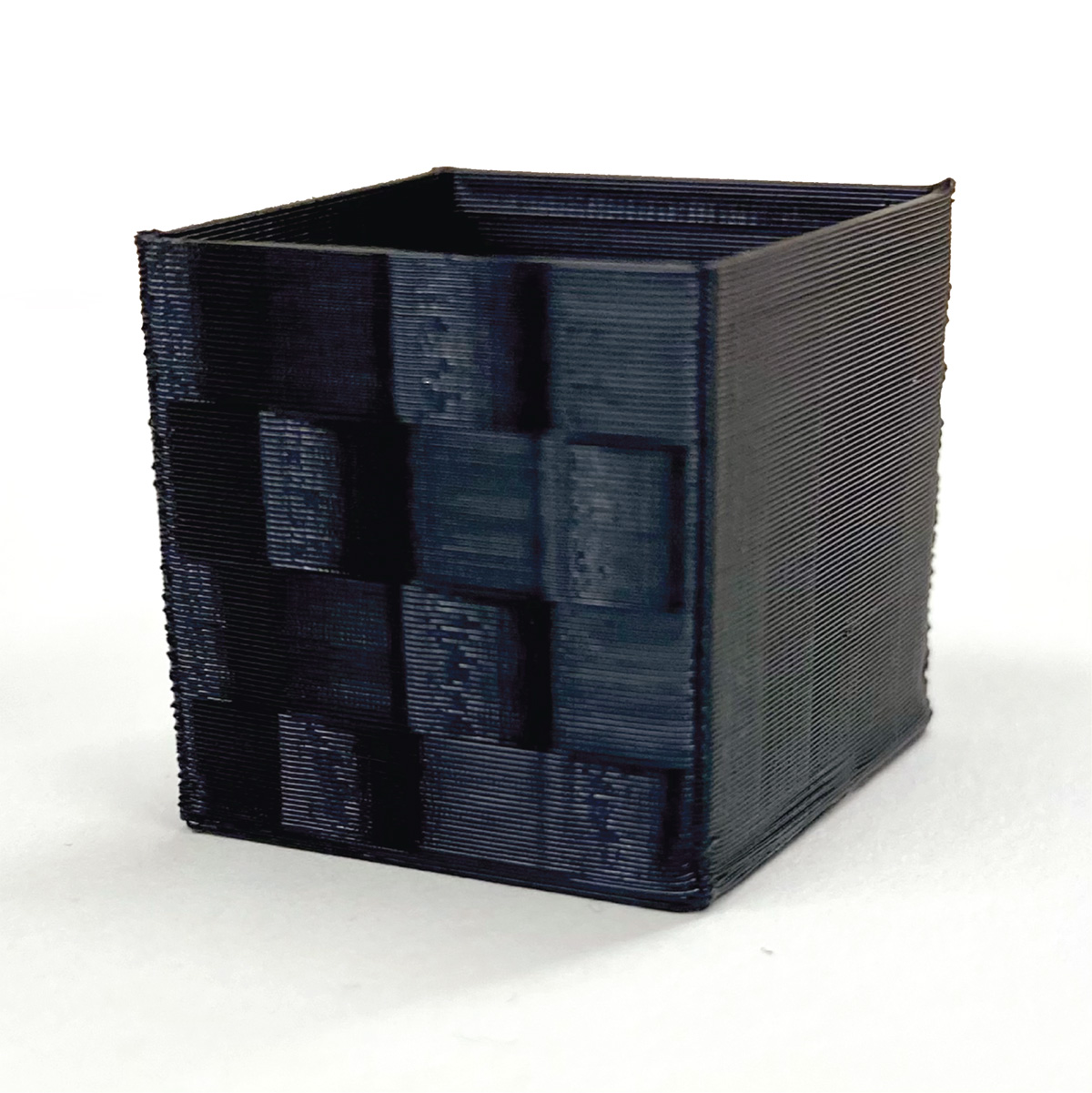}
\caption{The code on the left is varying the speed of extrusion along different regions of the cube, which results in different surface finishes in the final piece. On the right you can see the cube has a checkerboard pattern as a result of this ``velocity painting'' \cite{wheadon_velocity_2022} . With p5.fab, we can programmatically generate different textures.}
\label{fig:velocity-painting}
\end{figure}

Figure \ref{fig:velocity-painting} presents code for a 3-dimensional object and the resulting print. We use an approach called ``velocity painting'' to add a checkered pattern to one face of a cube \cite{wheadon_velocity_2022}. The technique creates different surface finishes by moving the printhead at different speeds. Velocity painting is normally achieved as a post-processing step: pixel data from a dithered image is applied to G-Code to modify print speeds. In \texttt{p5.fab}, we define the pattern directly by manipulating toolpath speeds. Programmatic texture generation offers an exciting opportunity to explore new surface finishes. These example prints demonstrate a range of use cases for \texttt{p5.fab}, as well as how a user might quickly iterate within a design space using a parametric design.

\subsection{System Implementation}
Our system includes the \texttt{p5.fab} library and a user interface which includes a toolpath visualization and GUI elements for sending commands and receiving positions. Figure \ref{fig:systemoverview} shows the interface.

\subsubsection*{Direct Communication} We communicate with the printer from a computer using serial communication over a wired USB connection. We leverage WebSerial support which allows us to read and write serial data from the browser. As of writing, WebSerial is supported in Google Chrome and Microsoft Edge, restricting use to these browsers. In exchange, we need only launch a local web server with no supplementary applications. Commands are added to a queue which are streamed to the printer. To ensure reliable communication, we await confirmation from the printer that the command has been added to the buffer before transmitting subsequent commands. While streaming, we can inject commands via the web console or custom user interface elements. These commands will be added to the top of the queue. 

Different machines will expect different commands depending on the firmware used by the control board. We develop \texttt{p5.fab} for use with Marlin firmware. In our survey, 57\% of respondents use Marlin. More respond that they are not sure and use the stock firmware installed on their printer. Almost half of respondents own a Creality Ender series which ships with Marlin by default. We therefore test our system on an unmodified Creality Ender-3 Pro. Our system requires no printer modifications---it is therefore relevant for a large number of machines and practitioners. \texttt{p5.fab} has also been extended to control the popular AxiDraw plotter.

\subsubsection*{Toolpath Visualization} The system offers a simple visualization of the toolpaths. The system will parse the G-code generated by the operator and render to the screen using p5. The visualization is shown in reference to the machine's work envelope, or the dimensional boundaries of the machine's operating area. We monitor system state such as current position and temperature in real-time as data is received from the printer. This data can be presented on screen or used as a variable in code to affect the toolpath. By designing and visualizing toolpaths directly, our system avoids dancing between separate CAD and CAM software.

\subsubsection*{Toolpath Manipulation} In popular FFF 3D printers, an extruder can move within the work envelope by specifying XYZ coordinates. The extruder axis, which extrudes filament through the nozzle, can also be set to a position. Our system offers control over these motion settings: for example, moving to position $(x, y, z)$ at speed $s$ while moving the extrusion axis to position $e$:  \lstinline{moveExtrude(x, y, z, s, e)}; setting the max acceleration for each coordinate axis: \lstinline{setMaxAcceleration(ax, ay, az, ae)}; and setting jerk for each coordinate axis: \lstinline{setJerk(jx, jy, jz, je)}. We also expose easy access to a number of other convenient G-code commands (e.g. \lstinline{autoHome}, \lstinline{setNozzleTemp}). Beyond built-in methods, we can send any arbitrary G-code to the machine. This level of toolpath control cannot be achieved in slicers which plan toolpaths based on geometric models. We forgo sophisticated toolpath planning in exchange for programmatic direct control. Some host applications offer a simple control panel to jog the printhead around the bed, send individual commands to the printer, and scaffold slightly more complicated behaviour such as pauses at specified layers. However, these features are for one-off use cases; by comparison, our system helps choreograph the large number of moves needed to print an entire model, access and procedurally manipulate higher order motion attributes, and consolidate bespoke multi-step workflows.

\begin{figure*}[htbp]
    \begin{minipage}{0.6\linewidth}
    \begin{lstlisting}[language=javascript]
fab.setStartingAcceleration(max(ax, ay, az));
fab.setMaxAcceleration(ax, ay, az);
for (x = xStart; x <= xEnd; x += xStep) {
  /* p5's map function remaps the value
     of the first argument from the range
     represented by arguments 2 and 3 to
     that of arguments 4 and 5. */
  theta = map(x, xStart, xEnd, 0, TWO_PI);
  z = A * cos(theta + PI) + A;
  fab.moveExtrude(x, y, z);
}
    \end{lstlisting}
    \end{minipage}
    \begin{minipage}{0.39\linewidth}
    \includegraphics[width=1\linewidth]{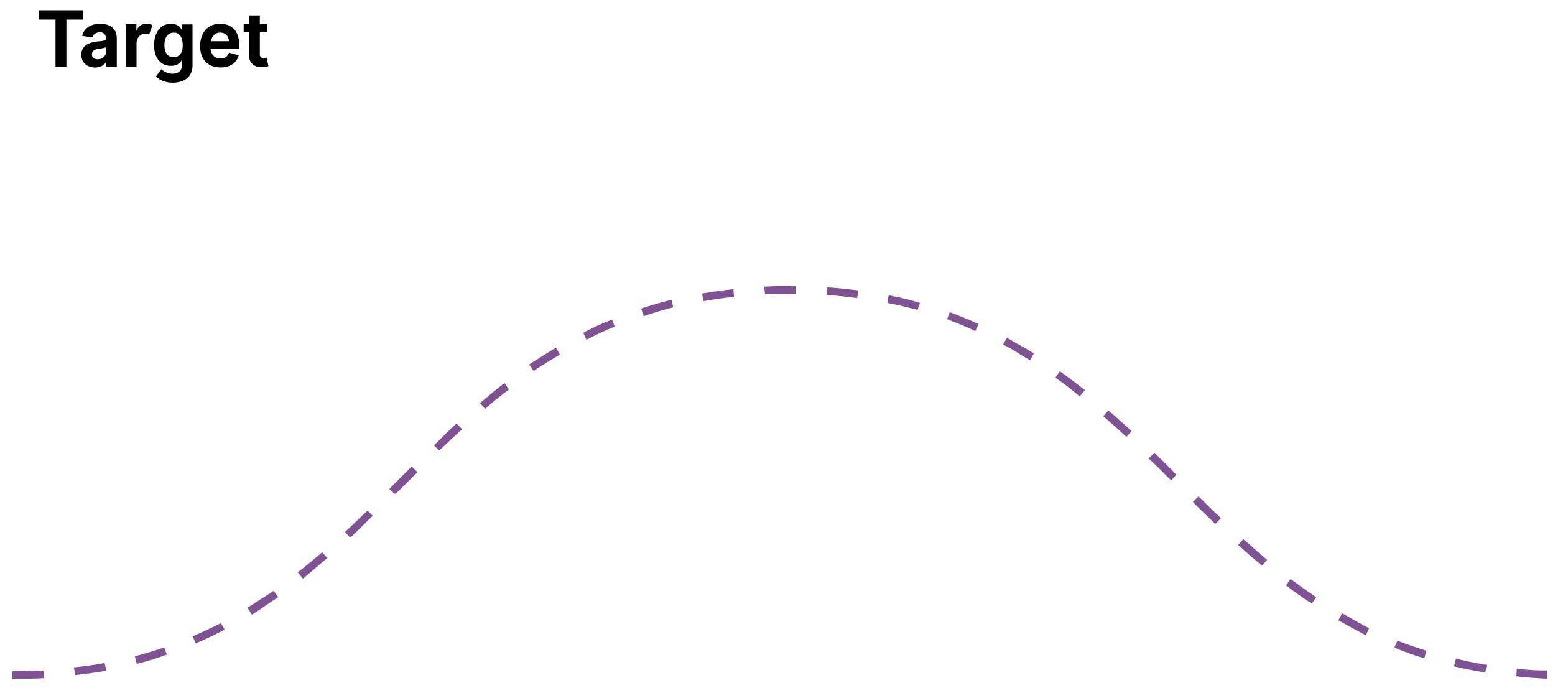}
    \end{minipage}
    \includegraphics[width=1\linewidth]{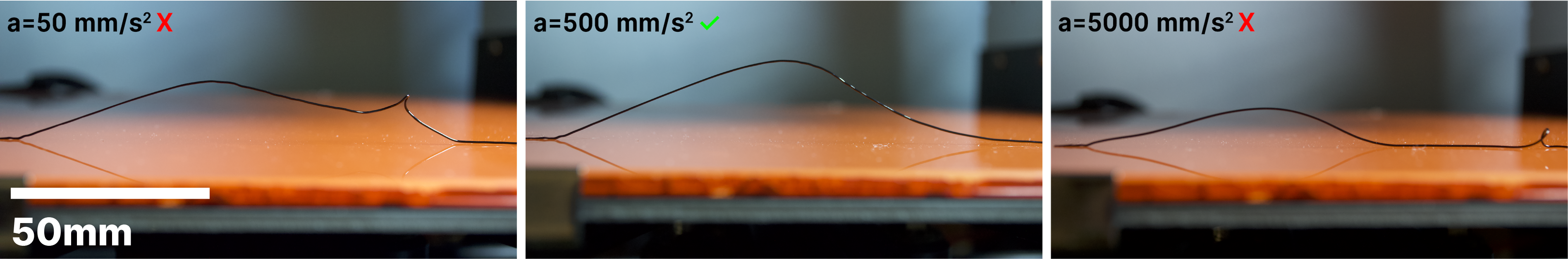}
    \caption{Tuning acceleration parameters. 3D printing without support structures can enable shorter fabrication times and different geometries. However, printing without support is prone to failure due to material collapse and needs to be fine-tuned for successful prints. To achieve the target form (top right), we test the effect of travel acceleration. The example code shown will move the printhead along a wave in the XZ plane. Acceleration for each axis can be controlled independently; output is shown for $a = ax = ay = az$. Bottom Left: The printhead obstructs the filament when accelerating slowly. Bottom right: The filament sags when accelerating quickly. Bottom center: closest approximation of the target form.}
    \label{fig:acceleration}
\end{figure*}

\subsection{Example Workflows}
\label{example-workflows}
We provide example workflows to demonstrate a range of system functionality including parameter tuning to generate experimental prints and a calibration routine to print on top of an existing object. We present each workflow alongside brief reflections with respect to our design goals.

\subsubsection*{Parameter Tuning}
Motivated by the tuning work described by participants in our formative study, we demonstrate \texttt{p5.fab} in the context of parameter tuning scenarios. The code in Figure \ref{fig:acceleration} generates a sinusoidal toolpath in the XZ plane. In typical slicers designing the toolpath illustrated would be nontrivial; slicers are not meant to extrude filament off of the XY plane. We tune acceleration parameters to achieve the target waveform. Fine-grain toolpath control reveals a range of material outcomes; at slower accelerations the printhead obstructs the filament, while at faster accelerations the filament sags.

Toolpath control also grants access to different methods for material deposition. Figure \ref{fig:dot-bridge} presents the scenario described in \ref{using-p5fab}. We over-extrude filament at an $(x,y)$ point while moving up in the $z$ direction to create ellipsoid dots in a single extrusion. This is accomplished by explicitly providing an extrusion quantity and velocity. Again, this behaviour is not possible to achieve with slicers which optimize toolpaths assuming traditional printing goals. A similar approach could be used for experimental 3D printing techniques such as z-pinning \cite{Duty2017}.
We illustrate a tuning process to connect these dots with a free hanging extrusion. At slower speeds the filament falls to the ground, while at higher speeds it is stretched. These structures can be used as complex `layers' to create the prints in Figures \ref{fig:teaser} and \ref{fig:experimentalprints}.

\subsubsection*{Experimental Prints}
\begin{figure*}[htbp]
    \begin{minipage}{0.6\linewidth}
    \begin{lstlisting}[language=javascript]
/* Print two dots while explicitly providing
  velocity and extrusion amount. */
fab.moveExtrude(xStart, y, z + dotHeight,
                slowSpeed, eAmount);
fab.moveRetract(xEnd, y, z);
fab.moveExtrude(xEnd, y, z + dotHeight,
                slowSpeed, eAmount);
  
/* Print a bridge with a quicker velocity. */
fab.moveExtrude(xStart, y, z + dotHeight, fastSpeed);
    \end{lstlisting}
    \end{minipage}
    \begin{minipage}{0.39\linewidth}
    \includegraphics[width=1\linewidth]{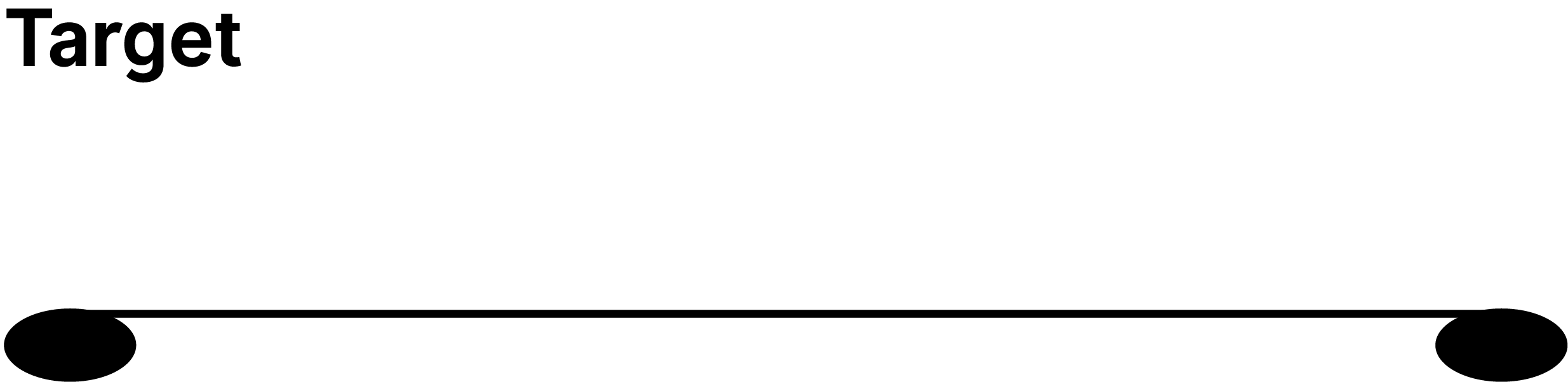}
    \end{minipage}
    \includegraphics[width=1\linewidth]{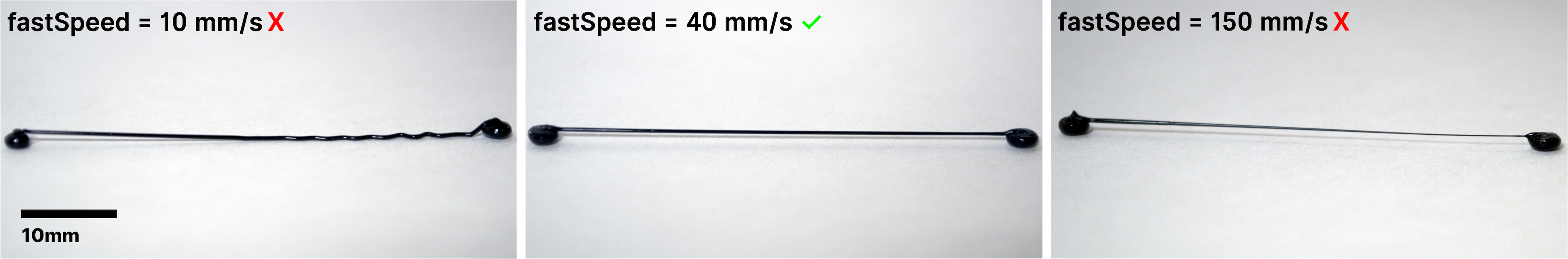}
    \caption{Tuning \lstinline{moveExtrude} parameters. Top right: We can over-extrude to create anchor dots connected by single line extrusion bridges. Bottom left: moving too slowly results in the filament falling. Bottom right: moving quickly stretches the filament. Bottom center: closest approximation of target form.}
    \label{fig:dot-bridge}
\end{figure*}

The fine-grain toolpath control which is required for tuning can be leveraged for expressive gain. We print a series of structures which explore toolpath geometry, motion, and extrusion. A collage of prints are shown in Figure \ref{fig:experimentalprints}. We reached each of these designs through a process of material exploration of various machine parameters. Figure \ref{fig:teaser} and Figure \ref{fig:experimentalprints} (left) show vases inspired by \textit{Filament Sculptures} \cite{lia_filament_2021}. The structures tuned in Figure \ref{fig:dot-bridge} become the building blocks of more complex assemblies.

Similarly, the sinusoidal toolpath from Figure \ref{fig:acceleration} is used to create prints with texture. Typical 3D printing will exhibit a `staircase effect' due to approximating slanted surfaces with parallel planes \cite{etienne_curvislicer_2019}. With curved toolpaths we can explore new surface finishes. Changing parameters produces structures which differentially respond to force. By tapping in to craft sensibilities fundamental to 3D printer calibration \& tuning practice, we encourage the practitioner to physically engage with the material and machine.

\subsubsection*{Unconventional Calibration}
Our formative study shows that manual bed leveling is a common calibration procedure. Many experienced practitioners view it as a necessary nuisance, and newcomers struggle to develop the requisite embodied skills---a task made harder when there is no process expert to support their learning. We synthesized these insights into a simple augmented bed leveling guide to send G-code commands in real-time to the printer.

\begin{figure}[h!]
    \centering
    \includegraphics[width=1\linewidth]{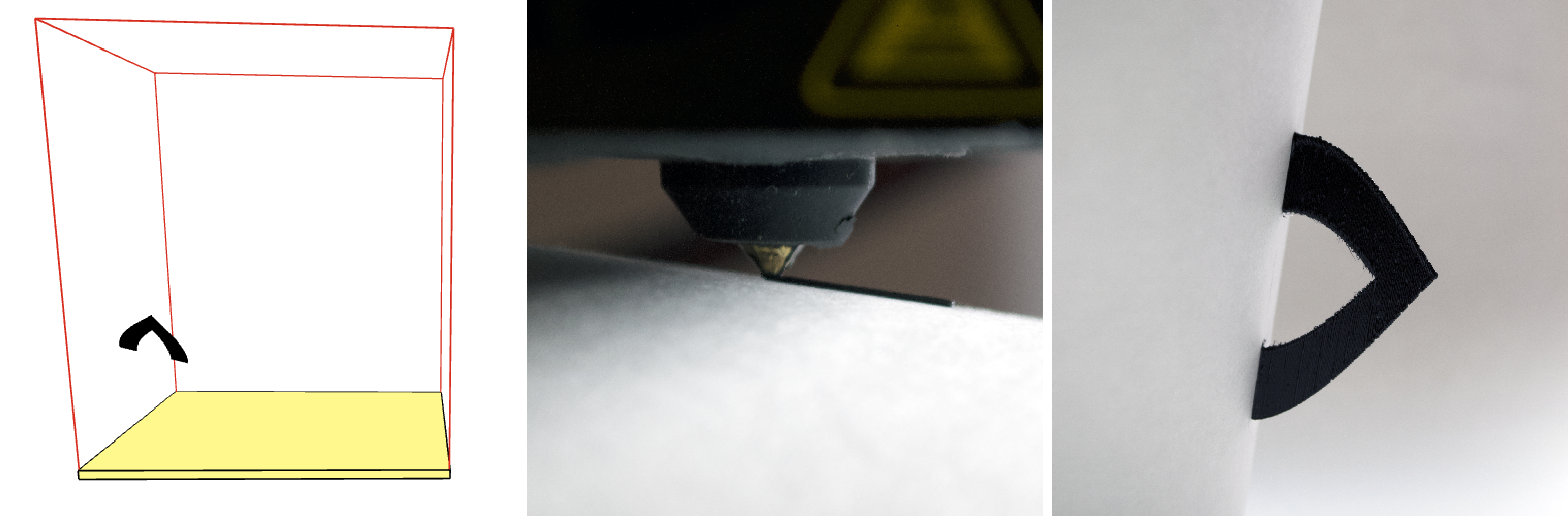}
    \caption{The \texttt{p5.fab} system makes it easy to print on top of existing objects, including ones that aren't flat. Shown here is a decorative handle being printed on a slanted cup. Left: after the operator aligns the nozzle to the desired points on the cup, the toolpath visualization shows the handle elevated from the print bed. Center: initial layer printing on the cup. Right: the resulting handle. }
    \label{fig:printcups}
\end{figure}

The system can be used for more unique interfaces in support of hybrid workflows. Take, for example, 3D printing directly onto an existing object which is elevated from the print bed. State of the art techniques involve novel machine augmentations involving 5-axis printing, milling, and scanning \cite{teibrich_patching_2015}. For everyday use, such fidelity might not be necessary. In traditional CAD/CAM software, printing on top of an existing object would require sophisticated coordination of measurements between the physical object, geometric model, and CAM settings. Iteration would prove difficult should these measurements not perfectly line up. This becomes more complicated still if the object in question is not also flat, parallel to the printer bed. We use \texttt{p5.fab} to build an interactive controls system to calibrate the printer for such a task. As an example, we design and print a handle onto a paper cup (Figure \ref{fig:printcups}).

The operator aligns the nozzle to the desired position on the cup and uses the controls interface to incrementally lower the nozzle until it touches a piece of paper. In this example, the operator chooses two points on the cup which define the handle endpoints. The precise geometry is generated on the fly, using the selected positions from the real world to set handle's length. Manually probing the endpoints accounts for the fact that handle's base is not flat; the side of the cup varies in the z-dimension. This also allows for quick iteration of geometry shape and size without returning to CAD software. If the print does not adhere well to the cup, it is straightforward to begin the calibration process again since no coordinates are hard-coded. Unconventional calibration offers low threshold opportunities for lo-fi object repair, patching, and hybrid aesthetics \cite{zoran_hybrid_2013}.

\section{Evaluating \texttt{p5.fab}}
Our process-driven insights suggest a productive role for calibration and tuning in creative practice. To more rigorously evaluate this connection, we conducted two workshops: the first with three artists working with code, and the second with three makers experienced with 3D printing. 

\subsection{Workshop Methodology}
In the first workshop, we recruited three professional artists (referred to as A1-3) who regularly use p5 in their work. Two of the participants had never used a 3D printer before, while the third had designed and printed various functional parts for art installations. The workshop lasted five hours. After an introduction to the tool, participants worked through three exercises. In analogy to the 2D sketching canvas offered by p5, participants created prints of single layer height. Then, they used a custom controls interface (introduced in \ref{example-workflows}) to perform a calibration routine to print a handle on top of a paper cup. Finally, the participants had time to explore the tool independently. We ran a complementary workshop with makers (referred to as M1-3) who reported advanced proficiency with FFF printing and beginner proficiency with coding. The workshop followed a similar structure, but the unconventional calibration exercise was substituted with an introduction to p5 and relevant programming concepts. Data collection consisted of participant observation, created artifacts, and written responses to survey questions collected after the end of each workshop.

This workshop evaluation strategy is limited in size and length. Given the limited time-frame we are focused on evaluating process and attitude rather than the quality of created artifacts. A longer-term study would provide a better sense of how \texttt{p5.fab} might integrate with practice. 

\subsection{Workshop Results}

\begin{figure*}[htbp]
    \centering
    \includegraphics[width=1\linewidth]{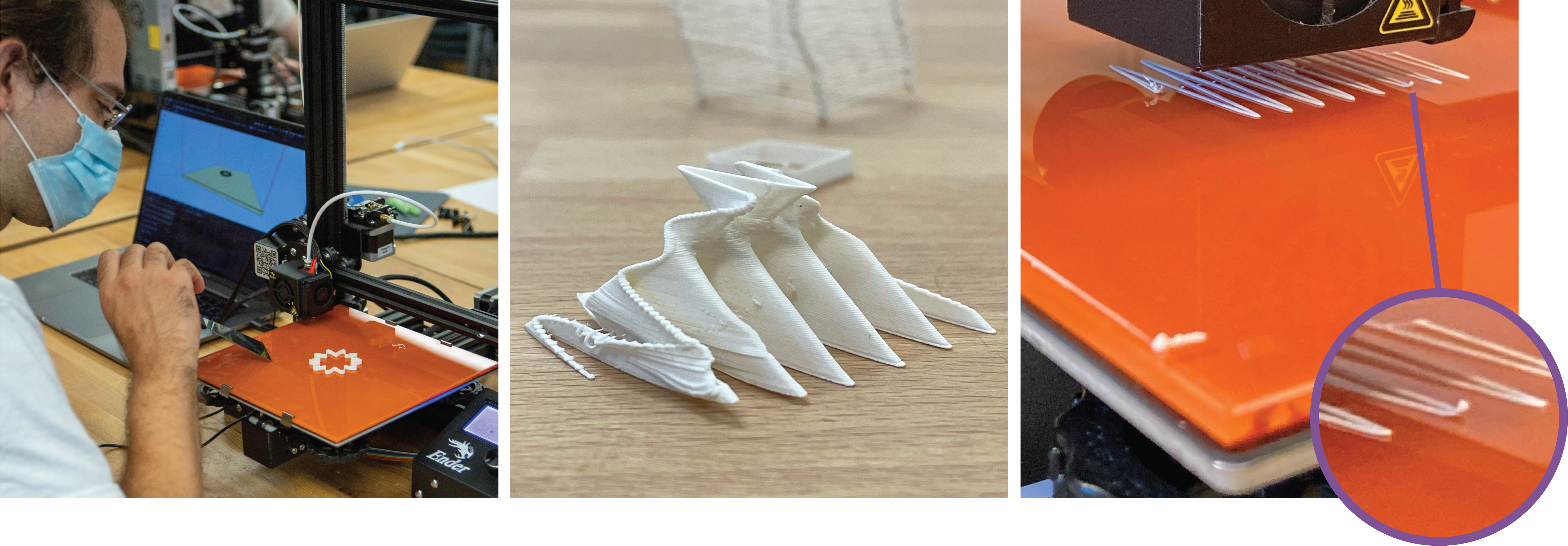}
    \caption{Process pictures from the artist workshop evaluation. Artists leveraged their creative coding skills to quickly begin printing (left). While printing a sinusoidal sculpture (center), A3 struggled with bed adhesion (right). To get the peaks and troughs to adhere, they negotiated their design geometry (wave period) with machine parameters (print speed).}
    \label{fig:workshopresults}
\end{figure*}

Participants in both workshops were able to begin printing with \texttt{p5.fab} in a matter of minutes. Throughout the workshops, participants explored the tool independently to create a range of artifacts and workflows. Here we recount strategies observed to iterate across digital and physical parameters. 

\subsubsection{Iterating on Digital Models}
For the artists, iterating on code generated models was a familiar activity. They voiced that learning to use \texttt{p5.fab} felt intuitive and accessible, even without having used a 3D printer. A major difference for the artists, however, was the time it takes to actually 3D print an object. A1 explains that in their usual creative coding workflow, they rapidly iterate. With \texttt{p5.fab}, there were bursts of rapid iteration with pauses during which they waited while something was being printed. As a result, the artists were keen to find ways to iterate faster. A3 kept a second `offline' window open which wasn't connected over serial to the printer. Here they would experiment with new forms using the toolpath visualization and move the code to their live session when their print finished. A2 scaled down their prints to get a sense of their visual outcome more quickly. 
 
The makers, on the other hand, found that streaming prints directly to the machine was much faster than their usual workflows. This was particularly true for small changes, where M3 explains that ``\textit{it's faster and easier to make quick changes to the code than having to remodel a section of the part, then slice, and print all over again}''. The ability to rapidly stop, start, and iterate on a design inspired M3 to use the printer bed as an accumulating 2D drawing canvas, adding more features to a scene including a stick figure and props. While small tweaks to an existing model were easily accomplished, M1 noted that that their minimal programming experience made it difficult to create new ones from scratch. M1 modeled linear forms and extrusions quickly but found it difficult to use code to describe forms which involved curves. This is in contrast to the artists who were comfortable generating surfaces described by mathematical functions. 
 
 Though coding proficiency determined the participants' design space, the system opened other novel opportunities at a low threshold. M2 was excited to print objects composed of fully three dimensional toolpaths: 
 \textit{``I’ve never printed anything non planar before! This has been a limitation on my mind for a while and this tool is a great introduction to getting past that barrier''}. Nonplanar here means that the toolpath is not constrained to the XY plane. M2 wrote parameterized models for cylindrical spirals to experiment with the relationship between layer height, speed, and radius.

Both artists and makers found that accounting for the materiality inherent to 3D printing required a carefulness in their coding process. Due to the time it takes to physically print an object, there was more at stake when modeling for fabrication. At one point, A3 incremented the z-dimension of their print before the y-dimension. This resulted in a failed print. Upon reflection, they noted that they had to pay particular attention to the order of their nested loops, something they normally don't think about when creating digital works. M2 had to ``\textit{think hard}" about when to move while extruding versus when to move while retracting to ensure no filament was deposited along transit moves in their model.

\begin{figure*}[htbp]
    \centering
    \includegraphics[width=1\linewidth]{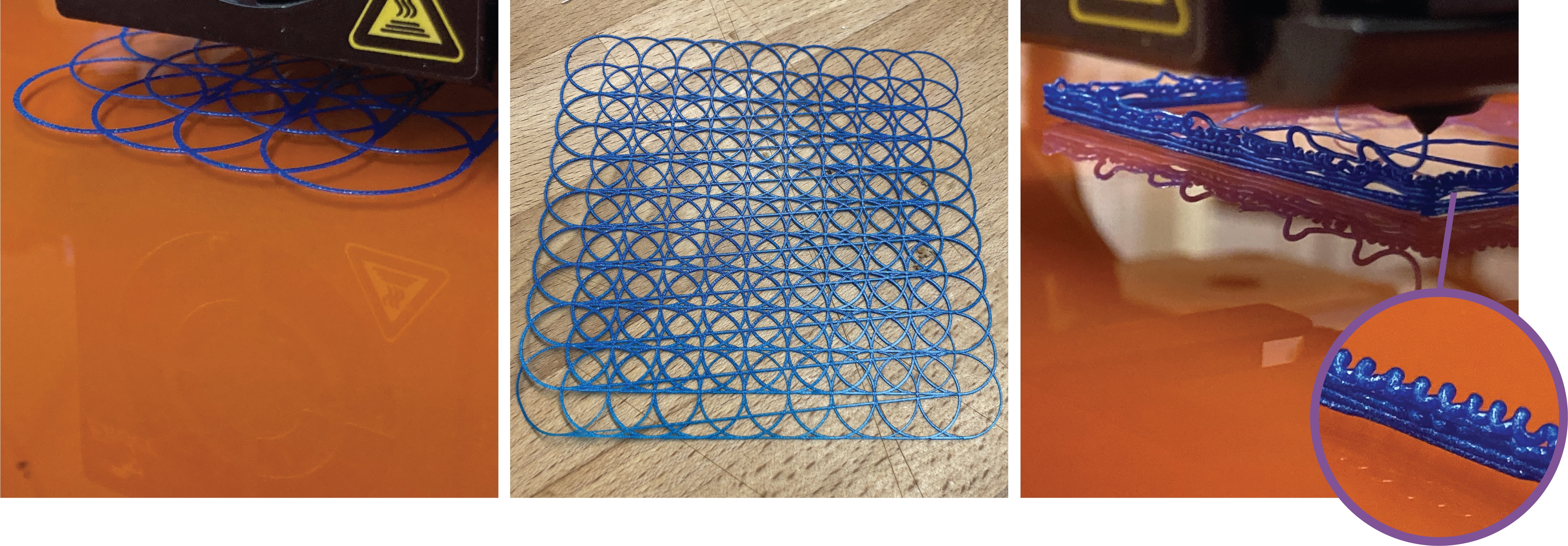}
    \caption{Process pictures from the maker workshop evaluation. After designing a model in code (left and center), makers were keen to tweak physical parameters to explore material behavior. M1 tried extruding differing amounts of filament from various heights to produce repeatable coils (right).}
    \label{fig:maker-workshop}
\end{figure*}

\subsubsection{Exploring Material-Machine Behaviour}
Early in the workshop, A3---who had some prior 3D printing experience---ran into trouble with first layer adhesion. While printing a structure with a base in the shape of a sine wave, the peaks and troughs refused to stick to the bed (Figure \ref{fig:workshopresults} right). In response, A3 began printing at a slower speed. When this did not completely resolve the issue, they decided to change the geometry. They found that increasing the period of the wave resulted in less extreme curvature which adhered better. Here we see toolpath control negotiating machine and model parameters simultaneously without reverting back to modeling software.

Artists with no 3D printing experience quickly encountered other common issues, like filament oozing from the nozzle. By tuning retraction settings they were able to minimize unwanted filament deposition. A2 considered using stringing as part of their design, reminiscent of prior HCI work \cite{laput_3d_2015}. Makers expected these issues and were able to handle them smoothly using the system. M1 made an interlaced single-layer structure similar to chain mail by overlapping extruded circles along a grid (Figure \ref{fig:maker-workshop}). This required excellent first-layer adhesion to avoid curling, and they noted that they were seated next to an open window which introduced wind and fluctuating temperatures that might affect adhesion. After a first failed print, they decreased their starting height in the z dimension instead of re-leveling the bed. Setting a z-offset is available on some auto-leveling printers and slicers, but this gave M1 a way to immediately account for their surroundings.

Coding in the context of 3D printing gave participants material feedback alongside standard debugging techniques. This was particularly the case for makers. M2 was trying to over-extrude filament but observed constant material deposition. After changing values drastically, they discovered that they misordered the function's arguments and were in fact changing speed. To explore the filament's material properties, M1 printed a cube using different extrusion multipliers on each side to deposit more or less filament (Figure \ref{fig:maker-workshop}). They began to hear a clicking noise from the printer which they recognized as the extrusion axis' stepper motor skipping as it unsuccessfully tried to keep pace. Finding these physical `error messages' guided M1's interaction with the system, who came to see \texttt{p5.fab} not as just a CAD alternative but as ``\textit{"a really immediate hands-on way to explore the weird edges of what the printer can do." }

The ability to explore and discover these ``weird edges'' is what excited our participants most across both workshops. In particular, A1 wrote that:
\begin{quote}
\textit{As someone who considers 3D printing to be a bit daunting (CAD software always seems like a huge barrier to entry), it was so liberating to just get direct control of the machine. The layers of abstraction that CAD/slicer software add are really boring to me, so not having to deal with that is so fun... Just give me the primitives!}
\end{quote}
Over the course of the workshop, A1 was increasingly drawn to the material properties of filament. Interested in what might not be printable with traditional tools, they experimented with extruding thick blobs of plastic because they appreciated the resulting organic shapes. They then tried to build these blobs into increasingly complex assemblies, including a 3 dimensional helix. They summarize their experience by saying that programmatic direct control \textit{``led me to investigate the expressive properties of the tool and material, rather than focusing on abstract shape geometry.''}, a success with respect to our design goals.

\section{Discussion}

\texttt{p5.fab} offers makers and creative coders a library to manipulate machine toolpaths. Below we discuss the results of our approach. We examine the importance of designing tools which prioritize material practice, the productive interplay between creative code and physical making, and possibilities for future systems.

\subsection{Tools to Support Material Practice}
In our workshops, participants found it useful that \texttt{p5.fab} prioritized the physical world. Artists with little to no 3D printing experience were able to build up mental models with \texttt{p5.fab}.  At first, A1 was not sure what speed is slow or fast for the printer, or how much filament is needed to under- or over-extrude. They were able to rapidly test values to build this intuition. They were therefore able to intuit toolpath changes to manage common printing problems and aesthetic concerns.

Even seasoned makers in our workshop expressed being ``\textit{overwhelmed}'' in the past by the amount of choices to make in a slicer. Issuing commands straight to the printer helped them discern ``\textit{the direct result of tweaking things like speed and extrusion}''. \citet{li_what_2021} found that in digital art tools, automation often added tedious manual labor instead of reducing it because artists lacked control over outcomes. Our formative study and workshops indicate a similar concern over lack of control in the toolpath planning process. Participants find it useful that \texttt{p5.fab} validates the manual tuning work intrinsic to digital fabrication practice. If desired, routines to automatically tune parameters could be expressed as functions written as code. Critically, the implementation of these routines requires initial experimentation via low level control, which \texttt{p5.fab} precisely provides. Time limitations of the workshop lead participants to focus on more dramatic attributes, like printhead velocity and total material deposition. Other physical considerations such as filament material and nozzle diameter, as well as higher order motion attributes like acceleration and jerk, provide a deeper catalog of possibilities.

Importantly, participants were able to apply knowledge gleaned from physical toolpath exploration towards new designs. In our workshops, participants printed novel forms and surface finishes. Projects such as \textit{Expressive FDM} have presented possibilities for novel aesthetics with 3D printers \cite{takahashi_expressive_2017}. Beyond identifying interesting behaviour, we find that it is critical to facilitate self-sufficient exploration. \texttt{p5.fab}'s programmatic approach to toolpath exploration offers a distinct way of interacting with machines. As a result, it extends the space of what is possible to make with digital fabrication tools.

\subsection{Bridging Creative Code and Physical Making}
Our example workflows and evaluation indicate productive intersections between creative code and physical making. Despite limited experience with 3D printing, the artists in our workshop picked up \texttt{p5.fab} smoothly. Artists learned from the ground up how the printer works. From their experiences, we are excited to bring the tool to existing creative coding communities as an alternative entry-point into digital fabrication. There are more opportunities to draw from creative code workflows in the design of \texttt{p5.fab}. Toolpath visualizations could update in real-time as code is adjusted, in analogy to live-coding practice. The system could also accommodate editing multiple versions of the current design without interrupting an ongoing print. These features can help encourage even faster digital iteration.

Makers had specific aspirations for 3D printing which they could not accomplish with their usual approaches. Participants came to the workshop wanting to make toolpaths not constrained the the XY plane and free-hanging extrusions. In large part, all participants were successful in these pursuits. To fully explore the possibilities, M3 reflected that they ``\textit{just need to work on getting better at controlling the process.}'' Without doubt, code comfortability played an important role in what participants were able to create. However, we recall from our formative study that CAD and CAM software also requires experience to achieve expertise. Tool building and sharing between coders and makers with complementary domain expertise is an exciting possibility of open-source tools, and we plan on building such community around ours. By doing so, we facilitate cross-pollination between makers and creative coders to explore a fuller range of relevant parameters, from digital models through physical material \& machine attributes.

\subsection{Towards Machine Interoperability}
Our design goals around material exploration, fine-grain control, and iterative practice suggest opportunities across more machines than we've covered here. \textit{Exquisite Fabrication} discusses how turn-taking between fabrication machines requires sensitizing oneself to the particularities of each new hybrid workflow \cite{goveia_da_rocha_exquisite_2021}. For example, work pieces need to be calibrated across work envelopes and operators must account for different machine tolerances. Our approach is well-suited to handle these concerns. \texttt{p5.fab} currently controls 3D printers running Marlin firmware and has been ported for use with the AxiDraw plotter. We plan on incorporating other FFF machines using common 3D printing firmware. We have also began to use \texttt{p5.fab} with the Potterbot clay printer. Ceramic 3D printing introduces a host of new concerns entirely different from FFF printing. Unlike spools of FFF compatible filament, clay requires manual preparation. Its material properties are not standardized which makes makes material exploration a critical concern. Controlling other fabrication tools such as laser cutters and desktop CNC mills introduce immediate opportunity for hybrid workflows. Moving to large format CNC mills will require considering the dependencies inherent in higher risk equipment. Ultimately, we hope p5.fab enables control not only in designing an individual artifact, but of heterogeneous workflows between machines and materials. 

\section{Conclusion}
We explored 3D printer calibration \& tuning practices as a way to inform the design of interfaces for digital fabrication machines.
The material practice of calibration and tuning is supported by our contribution \texttt{p5.fab}, a system to control machines from the creative coding environment p5.js. 
We use our system to explore 3D printing in new ways, such as designing toolpaths with tuned motion attributes and printing onto existing surfaces for hybrid workflows. In our evaluation with creative code artists and experienced makers, toolpath control helped participants negotiate common 3D printing problems with creative goals.
Given the size of the creative code community and increasing availability of personal fabrication machines, we believe \texttt{p5.fab} enables novel workflows which extend current opportunities in digital fabrication.

\section{Acknowledgements}
We would like to thank all of our interview and workshop participants for sharing their time and stories with us, including the following who wish to be named: Connor LeClaire, Rahul Banerjee, Clinton MacKinnon, Jackson King, Kyle Johnson, Kellie Dunne, Annie Liu, Alex Nagy, and Shelby Wilson. Many thanks to the p5.js contributors and community. Thanks also to Jasper Tran O'Leary for feedback on this work. This research is supported by the Alfred P. Sloan Foundation and the NSF through IIS Award 2007045.

\bibliographystyle{ACM-Reference-Format}
\bibliography{main}

\end{document}